\documentclass[12pt, draftclsnofoot, onecolumn]{IEEEtran}
\usepackage[T1]{fontenc}

\usepackage[shortcuts,acronym]{glossaries}

\usepackage{xcolor}

\usepackage{cite}

\ifCLASSINFOpdf
   \usepackage[pdftex]{graphicx}
  \usepackage{subcaption}
\else
\fi
\usepackage{amsmath}
\usepackage[cmintegrals]{newtxmath}
\usepackage[ruled]{algorithm2e}
\usepackage{caption}
\newcommand{\norm}[1]{\left\lVert#1\right\rVert}

\begin{document}
%
\title{AirNet: Neural Network Transmission \\ over the Air}
%
%
%

\author{Mikolaj Jankowski, Deniz G{\"u}nd{\"u}z and Krystian Mikolajczyk\\
{\tt\small \{mikolaj.jankowski17, d.gunduz, k.mikolajczyk\}@imperial.ac.uk}\\
Imperial College London
\thanks{The results in this paper were presented in part at the 2022 IEEE International Symposium on Information Theory (ISIT) \cite{airnet}.}}

\maketitle

\begin{abstract}
State-of-the-art performance for many edge applications is achieved by deep neural networks (DNNs). Often, these DNNs are location- and time-sensitive, and must be delivered over a wireless channel rapidly and efficiently. In this paper, we introduce AirNet, a family of novel training and transmission methods that allow DNNs to be efficiently delivered over wireless channels under stringent transmit power and latency constraints. This corresponds to a new class of joint source-channel coding problems, aimed at delivering DNNs with the goal of maximizing their accuracy at the receiver, rather than recovering them with high fidelity. In AirNet, we propose the direct mapping of the DNN parameters to transmitted channel symbols, while the network is trained to meet the channel constraints, and exhibit robustness against channel noise. AirNet achieves higher accuracy compared to separation-based alternatives. We further improve the performance of AirNet by pruning the network below the available bandwidth, and expanding it for improved robustness. We also benefit from unequal error protection by selectively expanding important layers of the network. Finally, we develop an approach, which simultaneously trains a spectrum of DNNs, each targeting a different channel condition, resolving the impractical memory requirements of training distinct networks for different channel conditions.
\end{abstract}


%

\IEEEpeerreviewmaketitle

\section{Introduction}


\IEEEPARstart{I}{n} recent years, deep learning (DL) has been shown to provide very promising solutions to many practical tasks within computer vision, natural language processing, robotics, autonomous driving, communications, and other fields. Developments within the area of DL have been made possible mainly thanks to the rapid growth of the computational power and memory available for both the researchers and the potential users of various DL-based algorithms. This resulted in the development of increasingly complex deep neural networks (DNNs) with millions and even billions of parameters trained on massive datasets, achieving impressive accuracy and performance in a wide variety of applications. On the other hand, the memory required to store a single modern DNN model can easily go from a few megabytes up to hundreds of gigabytes.

We typically evaluate the performance of a DNN architecture with the accuracy it achieves on new samples. This assumes the availability of the model at the end user. However, given the increasing prominence of DNNs employed for a large number and variety of tasks, we cannot expect every user to have all possible DNN parameters always available locally. Moreover, even a locally available model may need to be updated occasionally, either because the model at the server has been improved in the meantime through further training, or the task at hand has changed, e.g., due to variations in the statistics or size of the samples, or the system requirements. An alternative may be for the user to send its data samples to an edge server, where an up-to-date model is available for inference \cite{bottlenet, bn_plus_plus, jankowski_spawc, Kalor:SPAWC:21}. However, in many scenarios, the user may not want to send its data due to privacy constraints. Also, the user may want to infer many samples, which may create increased traffic. Moreover, the uplink capacity may be limited compared to the downlink. In such scenarios, a reasonable solution is to transmit the DNN parameters to the user over the network, rather than the user sending the data samples to the edge server. However, given the growing size of modern DNNs, and stringent latency requirements of edge intelligence applications,
the transmission of the DNN parameters to an edge user may be infeasible. This problem can be further amplified in the future by the adoption of very specialized DNNs, that either solve very specific tasks adapted to a specific geographic location, or are frequently updated due to the non-stationarity of the underlying tasks. In such scenarios, it is necessary to develop methods, which allow for fast and reliable delivery of DNN parameters over the wireless channel.

A fundamental ambition of the sixth generation (6G) of mobile wireless networks will be to enable seamless and ambient edge intelligence. Therefore, it is expected that the efficient storage and delivery of DNN parameters will constitute a significant amount of traffic. Indeed, model distribution and sharing for machine learning applications at the edge is already being considered by 3GPP as part of the next generation of mobile networks \cite{3gpp}. Our goal in this paper is to develop efficient DNN delivery techniques at the wireless network edge, such that the highest performance can be achieved by the user despite wireless channel imperfections.

\begin{figure}[t]
\begin{center}
\includegraphics[width=0.9\linewidth]{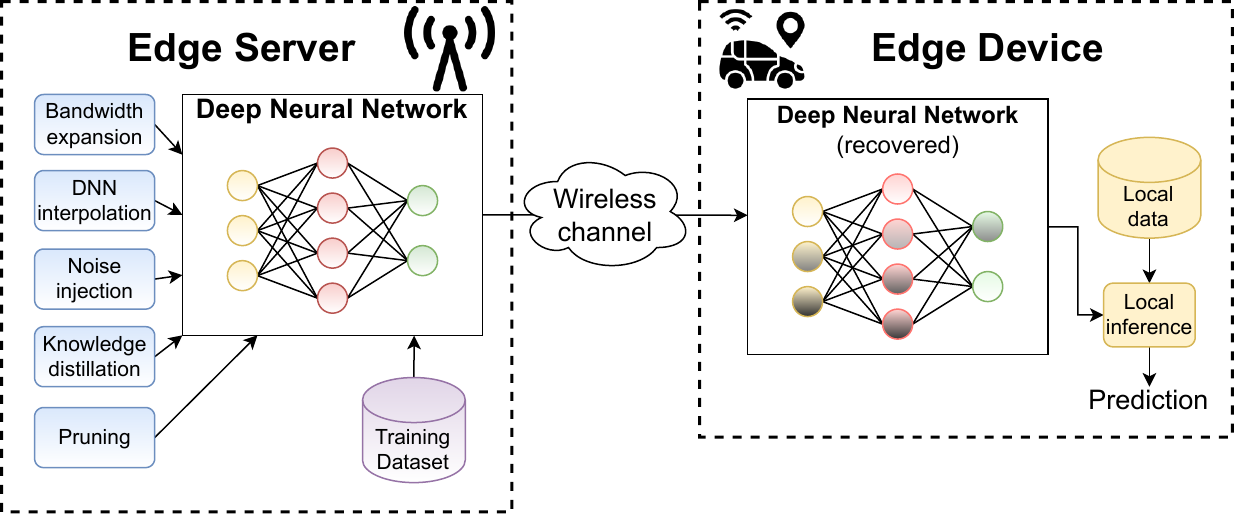}
\end{center}
  \caption{System model. In AirNet, DNN is transmitted over a wireless channel in an uncoded fashion, and it employs various training techniques aimed at bandwidth reduction and enabling robustness against channel noise. 
  }
\label{fig:system_model}
\end{figure}

We consider the system model illustrated in Fig. \ref{fig:system_model}, where an \textit{edge server}, e.g., a base station (BS), should enable an \textit{edge device}, e.g., a mobile phone, an autonomous car, a drone, a medical device, to carry out inference on local data samples. We impose a strict latency constraint as well as the usual resource constraints on the wireless channel between the server and device. The edge device reconstructs a local model to be used for inference on local samples. 

Consider, for example, vision- or LIDAR-aided channel estimation or beam selection, where an autonomous car aims at establishing a high-rate millimeter wave connection with a BS in the non-line-of-sight setting, based on the input from its cameras or LIDAR sensors \cite{mashhadi2021federated, mmwave_itu}. The best approach would be to provide the car with a DNN, optimized specifically for the coverage area of the particular BS. However, we cannot expect each car to store DNNs trained for every possible cell area that it may go through. Instead, it is much more reasonable to assume  that locally-optimized DNNs would be delivered to vehicles as they move around, and come into a coverage area of a particular cell.
Another important application of model delivery is federated/distributed learning over wireless channels \cite{gunduz_magazine, chen_distributed_learning}. In such problems, a locally trained/updated model is shared with a parameter server or neighboring devices at each iteration of the training process, and highly efficient delivery is essential considering that training a large DNN model can require thousands of iterations. The same would hold for many other DNN-aided edge applications that may require localized optimization of DNN parameters, e.g., various localization services. On the other hand, sending even a relatively simple VGG16 \cite{vgg} network requires transmission of roughly $15\times 10^6$ 32-bit floating-point parameters. Assuming a standard LTE connection at a channel signal-to-noise ratio (SNR) of $5\textrm{dB}$, and capacity-achieving channel codes, such a transmission would require roughly 30 seconds to complete, which is unacceptable for most time-sensitive edge applications.

In this work, we consider two fundamental approaches to this problem \cite{airnet}. As in most wireless lossy source delivery problems, the two approaches follow the separation approach and the joint source-channel coding (JSCC) approach, respectively. In the separation-based approach, we train a model of a certain size, whose parameters are then transmitted over the channel using an error correction code to provide reliability in the presence of noise and fading. We can either train a sufficiently small-size network that can be delivered over the channel, or a pre-trained network can be compressed to be reduced to the required size. In the alternative JSCC approach, the parameters of the network are transmitted directly over the channel. In the latter approach, called AirNet, the training is carried out taking into account the effect of channel imperfections. We remark that, in both approaches, we assume the availability of the data at the edge server, which allows to optimize the DNN in terms of accuracy, size, and robustness. The problem we face can be treated as a lossy source delivery problem over a wireless channel; more specifically a remote source delivery problem, where the goal is to deliver the underlying true inference function, i.e., the model, to the edge user. But, the edge server does not have access to the true model, and it can only estimate it through its local dataset.

Our major contributions can be summarized as follows:

\begin{itemize}
    \item We introduce a novel communication problem that requires the delivery of machine learning models over wireless links under strict bandwidth and transmission power constraints for reliable inference at the receiver. 
    \item We propose a novel JSCC approach to this problem, called AirNet, that can achieve reliable edge inference at very low channel SNR and bandwidth, and is robust to channel variations, as opposed to separation-based techniques, which break down abruptly when the channel SNR cannot support the adopted channel code rate.
    \item We use network pruning to meet the channel bandwidth constraint, and knowledge distillation (KD) to increase the accuracy of inference at the receiver. To further increase AirNet's robustness to adverse wireless channel conditions, we employ noise injection during training and carefully study its effect on performance.
    \item In order to provide unequal error protection (UEP) to different network layers, we employ bandwidth expansion; that is, we prune the network to a size smaller than the available bandwidth, and expand some of the layers to provide extra protection against channel noise. We choose the layers to be expanded by their \textit{sensitivity}, measured through the Hessian matrix. We show that UEP through bandwidth expansion provides significant gains in terms of the final accuracy at the receiver.
    \item Above algorithms are trained for a specific channel SNR to obtain the best accuracy. This would require training and storing a different set of network parameters for different channel conditions, which is not practical. To resolve this critical limitation, we propose an ensemble learning approach, where we obtain a spectrum of networks simultaneously for a whole range of channel SNRs.
    \item We present extensive evaluations of AirNet, including different datasets, channel models, training and pruning strategies, channel conditions, and power allocation methods. We show that the proposed AirNet architecture and training strategies achieve superior accuracy compared to separation-based methods, which employ DNN compression followed by separate channel coding. AirNet allows for a significant reduction in bandwidth requirements while sustaining satisfactory levels of accuracy for the delivered DNNs.
\end{itemize}

The remainder of the manuscript is organized as follows. We present relevant works on DNN compression and JSCC literature in Section \ref{sec:related_work}. In Section \ref{sec:system_model}, we present our system model, followed by the introduction of the AirNet architecture in Section \ref{sec:airnet} with all the details regarding training, pruning, and noise injection methods used. Section \ref{sec:uep} presents bandwidth expansion methods, alongside an UEP scheme, and Section \ref{sec:snr_robustness} introduces an ensemble learning scheme, which trains a spectrum of networks aimed at different values of channels SNRs. This is followed by Section \ref{sec:results}, which evaluates AirNet on various datasets, channel conditions and choices of training parameters. Finally, Section \ref{sec:conclusions} concludes the paper. 

\section{Related Work}
\label{sec:related_work}

\subsection{Network compression}

Despite achieving state-of-the-art results in many emerging machine learning tasks, DNNs usually require significant computation and memory resources. This causes an important problem at the wireless network edge: new and larger DNNs are proposed each year, which frequently need to be distributed across networks. However, DNNs are usually overparameterized, thus their complexity can be reduced significantly without sacrificing accuracy.

\textbf{Network pruning.\indent} Pruning DNNs has been originally proposed in \cite{optimal_brain_damage}. In recent years, many pruning algorithms have been proposed \cite{pruning_molochanov, pruning_molochanov2, l1_pruning, lasso_pruning, snip_pruning, variational_pruning, thinet, lottery_ticket_pruning, hrank_pruning, succesive_pruning}. The main difference between pruning algorithms is the saliency measure used to determine the importance of each parameter. In \cite{pruning_molochanov, pruning_molochanov2}, Taylor expansion is used to approximate the change in the loss function induced by pruning. L1-norm of the network weights is considered in \cite{l1_pruning}. More complex methods of selecting the most important convolutional filters are proposed in \cite{lasso_pruning, variational_pruning, hrank_pruning}. Authors of \cite{snip_pruning} consider a single-shot network pruning strategy. We note, that majority of the state-of-the-art pruning algorithms perform sequential pruning and fine-tuning steps to regain the accuracy loss induced by removing certain parameters. Findings of \cite{lottery_ticket_pruning} seem to confirm that this is a proper approach, since finding smaller DNNs and training them from scratch usually leads to a sub-optimal performance.

\textbf{Network quantization.\indent} Another method of reducing the complexity of DNNs is quantization. Instead of using a full, 32-bit precision for storing the network weights and activations, low-bit precision can be used, resulting in significant gains in terms of both the computations and the memory footprint. Many works have studied network quantization in recent years \cite{quantization_networks, 8bit_quantization, hawq, zeroq_quantization, lqnets_quantization, adabits_quantization, post_training_quantization}. These works study different aspects of quantization, e.g., evaluation of the sensitivity of the DNN parameters, training strategies that benefit quantization, etc. Authors of \cite{hawq} estimate the statistics of the Hessian matrix corresponding to each layer of the network, in order to derive a layer-dependent sensitivity metric for a mixed-precision quantization process. In the DeepCABAC method \cite{deepCABAC}, quantized DNN parameters are further compressed by utilizing context-adaptive binary arithmetic coding.

Analog storage of network parameters is studied in \cite{compression_for_storage_devices}, where the authors also consider applying channel noise to DNN parameters during training, pruning, and KD. We note that despite some techniques can be effectively used for both analog storage and wireless transmission of DNNs parameters, there are many differences. The fundamental difference between these two applications is that for analog storage, the channel noise variance does not change with time, but rather with the magnitude of the stored value. For wireless transmission, however, we have to ensure that the network adapts well to a variety of channel SNRs. To this end, our approach requires either storage of multiple DNNs, each trained for a specific value of channel SNR, or a training method, which ensures that the network can adapt to various SNRs. Another novelty of our work is that we take into consideration the sensitivity of each layer with respect to channel noise, and assign more channel bandwidth to the most sensitive layers.

\subsection{Deep JSCC}
As we have highlighted above, the considered DNN delivery problem is a JSCC problem. Although Shannon's separation theorem \cite{shannon} dictates the optimality of separate source and channel coding, it holds under the assumptions of infinite source and channel bandwidths, ergodic source and channel distributions, and for an additive distortion measure in general, all of which are violated in our problem.
More recently, DNN-based efficient JSCC techniques have been shown to outperform conventional digital approaches even for the wireless transmission of well-studied sources such as images \cite{jscc_image_dnn, jscc_feedback, Kurka:TWC:21, Yang:ICC:21, Wu:WCL:22, dai2022nonlinear}, speech \cite{Weng:JSAC:21}, or videos \cite{roy_video}.
The approach proposed in \cite{jscc_image_dnn}, called DeepJSCC, consists of end-to-end training of an autoencoder network with a communication channel model embedded into the architecture, between the encoder and the decoder. The authors propose to directly map input image pixel values to real or complex-valued channel symbols, and show that JSCC outperforms standard compressive codecs (BPG, JPEG2000) concatenated with state-of-the-art channel codes (LDPC).
DeepJSCC has also been applied to other downstream tasks, such as remote classification \cite{bn_plus_plus, jankowski_spawc}, retrieval \cite{jankowski_JSAC}, or anomaly detection \cite{Kalor:SPAWC:21} problems.
Although the problem of DNN delivery over wireless channels is a JSCC problem at the core, it is substantially different from the delivery of other common sources, such as image or video. Not only it has a very different measure of quality at the receiver, but also, unlike typical information sources, we do not have a dataset of DNNs with common statistics at the transmitted, that can be exploited for efficient compression or JSCC. Instead, with the training data available, a particular DNN architecture can be trained or fine-tuned specifically for efficient wireless delivery. A similar problem involving wireless delivery of DNN parameters is studied in \cite{yulin_dnn}, but in the absence of training data.

\section{System Model}
\label{sec:system_model}

We consider an edge server, which has access to a labeled dataset that can be used to train a machine learning model. However, this trained model is to be employed at an edge device to predict labels of new data samples. The edge server is capable of training a model locally, but it is connected to the edge device through a bandwidth and power-limited noisy channel. The goal is to minimize the performance loss, measured by the accuracy of the model at the receiver end, due to the channel imperfections, while meeting the prescribed bandwidth and power constraints at the transmitter. This is different from conventional channel coding problems, which aim at minimizing the probability of error, or the conventional JSCC problems, which minimize an additive distortion measure defined on source samples. The optimal performance of such separation-based or JSCC approaches is usually achieved under the presence of multiple samples from the source distribution, which allows generating a model of a source used for compression and reconstruction with minimal error/distortion. In our case, only a few, or a single DNN is stored at the edge server, which cannot obtain such a source model. AirNet is still a JSCC problem, but with an unconventional distortion measure, that requires not only training but also delivery of a model over a noisy channel with good generalizability properties. 

While the above formulation is general enough to be applied to any learning model, given their state-of-the-art performance and large size that require significant communication resources, we focus on DNNs. More specifically, we consider a DNN with parameters $\mathbf{w} \in \mathbb{R}^d$, trained at the edge server. The DNN is then transmitted to the edge device over the wireless channel. The specific channel models used in our work are described in Section \ref{subs:channel_model}. After the transmission, the edge device reconstruct another network $\tilde{\mathbf{w}} \in \mathbb{R}^d$ based on the signal it receives over the channel. This network is then employed at the device to obtain predictions $p_{\tilde{\mathbf{w}}} = f_{\tilde{\mathbf{w}}}(I)$, where $I$ is a sample from the edge device's local dataset, and $f_{\tilde{\mathbf{w}}}(\cdot)$ represents the forward pass of the DNN parameterized by the decoded weights $\tilde{\mathbf{w}}$.

\subsection{Channel model}
\label{subs:channel_model}
We model the channel between the edge device and the edge server as an additive white Gaussian noise (AWGN) channel. We consider static as well as slow fading channels. For the static AWGN channel, we have $\mathbf{y} = \mathbf{x} + \mathbf{z}$, where $\mathbf{x} \in \mathbb{C}^b$ is the channel input with the channel bandwidth $b$, defined as the number of channel uses, $\mathbf{y} \in \mathbb{C}^b$ is the channel output, and $\mathbf{z} \in \mathbb{C}^b$ is a vector containing independent and identically distributed (i.i.d.) noise samples drawn from circularly-symmetric complex Gaussian distribution $\mathcal{C}\mathcal{N}(0, \sigma^2)$ with variance $\sigma^2$. An average power constraint is imposed on the channel input, i.e., $\frac{1}{b} \sum_{i=1}^b \norm{x_i}^2 \leq P$. We set $P=1$ without loss of generality, which corresponds to an SNR of $\mathrm{SNR} = 10\log_{10}\left(\frac{1}{\sigma^2}\right)$.

In the slow fading scenario, the channel output $\mathbf{y} \in \mathbb{C}^b$ is given by $\mathbf{y} = h\mathbf{x} + \mathbf{z}$, where $h\sim \mathcal{C}\mathcal{N}(0, \sigma^2_h)$ is the channel gain. We assume that the channel remains constant for the duration of a block of $b$ channel symbols, but takes i.i.d. values drawn from $\mathcal{C}\mathcal{N}(0, \sigma_h^2)$ across different blocks. As in the static channel scenario, we impose the average power constraint of $P=1$, and the average SNR is given by $\mathrm{SNR} = 10\log_{10}\left(\frac{\sigma^2_h}{\sigma^2}\right)$. In all our experiments we set $\sigma_h$ to $1$. We also assume perfect channel state information (CSI) to be available at the receiver, thus, to get rid of the multiplicative component $h$, the receiver scales received signal $\mathbf{y}$ by $\frac{h^*}{\norm{h}}$. The resulting signal is given by $\mathbf{x} + \frac{h^*\mathbf{z}}{\norm{h}^2}$, which is equivalent to AWGN channel with a random SNR value. Therefore, in the case of communication over fading channels, we need to come up with a transmission scheme that will perform well over a range of SNRs with a given average variance.

\section{AirNet: JSCC of DNN Parameters for Reliable Edge Inference}
\label{sec:airnet}

The conventional approach to the problem presented above would be to train a DNN of limited size, and quantize, compress and deliver its parameters over the channel using state-of-the-art channel codes. While we will also consider this ``separation-based'' approach as a baseline, our main contribution is a JSCC approach, where the DNN parameters are directly mapped to channel inputs. Next, we present the details of this approach.

\subsection{Training strategy}
Our performance measure is the average accuracy of the model reconstructed at the edge device on new samples. Here, the randomness stems from not only the random and previously unseen samples encountered at the receiver, but also from the channel noise and fading.

We first train a DNN with the data available at the edge server. In the proposed AirNet approach, each DNN parameter will be mapped to a channel input symbol. This has two consequences: first, the number of DNN parameters that can be delivered is limited by the channel bandwidth $b$; and second, transmitted parameters are received with random noise at the receiver. To increase network's robustness against noise we inject a certain amount of noise to the network's weights during training.
While we initially train a large DNN with more than $b$ parameters, in the second training step, we prune them by removing redundant parameters in order to satisfy the bandwidth constraint.
Alongside pruning and noise injection, we also apply KD to prevent accuracy drop due to pruning.
We provide the details of each training step in the rest of this section. 

\subsection{Pruning}
\label{subs:pruning}
To reduce the bandwidth required to transmit the network parameters we apply a simple pruning strategy \cite{l1_pruning}. Pruning removes redundant parameters while maintaining satisfactory performance, which effectively reduces the bandwidth requirement. In this work, we apply repetitive pruning and fine-tuning steps. At each pruning iteration, we remove a fraction of parameters that have the lowest $l_1$-norm. In order to avoid the transmission of meta-data containing the network's structure after pruning, we only consider structured pruning, which removes either entire convolutional filters or entire neurons, depending on the layer type. During fine-tuning, we simply re-train the network to recover the performance lost due to pruning. We additionally apply noise injection and KD to further increase robustness to channel noise and reduce the performance loss imposed by pruning.

\subsection{Noise injection}
\label{subs:noise_injection}
Noise injection has been originally proposed as a regularization method to prevent overfitting in DNNs \cite{noise_injection}. In this work, however, we utilize noise injection as a method to increase the DNN's robustness to channel noise. We hypothesize that, if the network experiences a certain amount of noise injected to its weights during training, it will effectively learn to achieve good performance even after its weights are transmitted over a noisy channel. At each iteration of the training, we apply the same noise components as imposed by the channel on the current network parameters, and calculate the loss function using these noisy network parameters. Please note we only inject the noise during network training to mimic the channel noise experienced by the network during the transmission.

\subsection{Knowledge distillation (KD)}
\label{subs:knowledge_distillation}
KD has been proposed as an effective method to boost up the performance of various DNN models trained for classification task \cite{knowledge_distillation}. In KD, a large DNN, called the \textit{teacher}, which achieves high accuracy in the task, distills some knowledge into a smaller DNN, called the \textit{student}. The loss function in KD is defined as a sum of two terms:

\begin{equation}
\label{eq:knowledge_distillation}
    L_{total} = -t^2 \sum_i^N \hat{p}_i \log p_i - \sum_i^N \bar{p}_i \log p_i,\text{ where }\hat{p}_i \triangleq \frac{e^{\frac{\tilde{p}_i}{t}}}{\sum_j^N e^{\frac{\tilde{p}_j}{t}}}.
\end{equation}
In Eq. (\ref{eq:knowledge_distillation}), the first term is responsible for distilling the knowledge between the teacher and student, and the second is a standard cross-entropy loss, where $\hat{p}_i$ represents the softmax predictions of the teacher model, $t$ is the temperature parameter, set to $2$ in all our experiments, $\bar{p}_i$ denote the ground truth, and $p_i$ are the student's predictions. 

\section{AirNet with Unequal Error Protection (UEP)}
\label{sec:uep}
\label{subs:bandwidth_expansion}

The network trained with noise injection has a certain level of robustness against channel noise. However, the performance will degrade inevitably as the SNR decreases. Here, we first propose trading-off the pruned network size with robustness against noise. The main idea is to use bandwidth expansion to better protect the DNN parameters against noise. For example, instead of pruning the network down to $b$ parameters, we can prune it to, say, $b/2$ parameters, and use two channel symbols to transmit each network parameter. We consider two bandwidth expansion methods: Shannon-Kotelnikov (SK) mapping \cite{kotelnikov} and simple \textit{layer repetition}. 

SK mapping has been successfully used in JSCC in \cite{archimedes_coding}, where the authors use Archimedes' spiral as a codebook, and show its benefits for both bandwidth compression and expansion tasks. In this work, we employ a similar approach. Specifically, we map the DNN parameters onto Archimedes' spirals defined as:

\begin{equation}
\label{eq:spiral1}
    x_1 = \frac{\Delta}{\pi} w \cos (\gamma w),\ x_2 = \frac{\Delta}{\pi} w \sin (\gamma w), w \geq 0
\end{equation}
\begin{equation}
    x_1 = - \frac{\Delta}{\pi} w \cos (-\gamma w + \pi),\ x_2 = - \frac{\Delta}{\pi} w \sin (-\gamma w + \pi), w<0,
    \label{eq:spiral2}
\end{equation}
where $\Delta$ is a scaling factor, which we set to $1$, $\gamma$ controls the length of the spirals without changing the radius of the disc occupied by the spiral, and $w$ is a DNN parameter.

\begin{figure}[t]
    \centering
    \begin{subfigure}[b]{.49\textwidth}
        \centering
        \includegraphics[width=\textwidth]{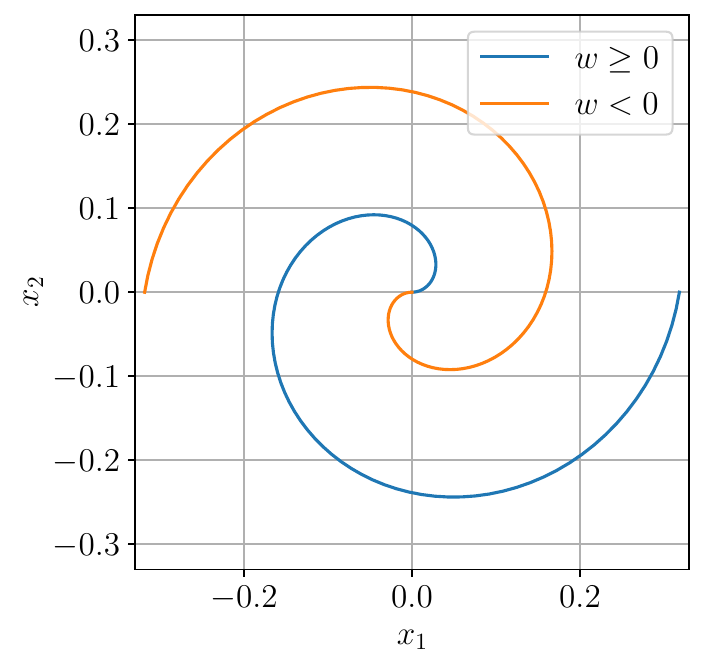}
        \caption{$\gamma = 2\pi$}
        \label{fig:archimedes_2pi}
    \end{subfigure}
    \hfill
    \begin{subfigure}[b]{.49\textwidth}
        \centering
        \includegraphics[width=\textwidth]{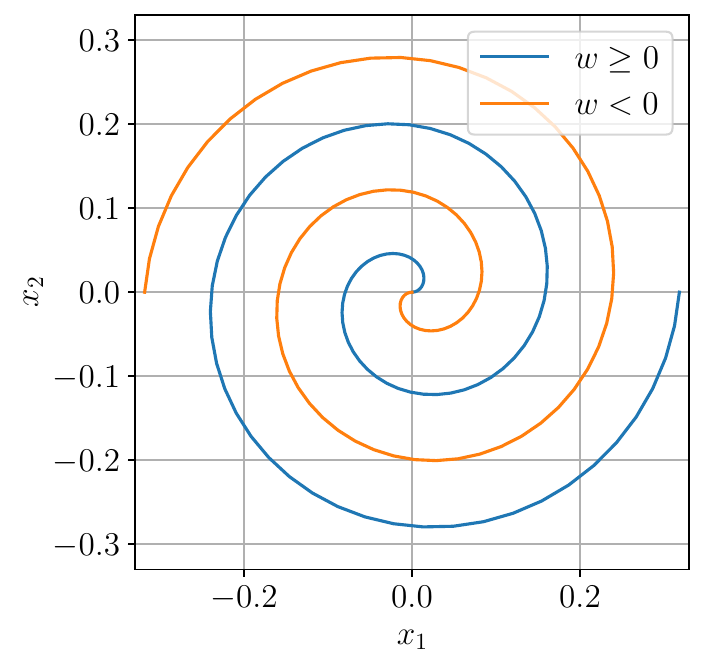}
        \caption{$\gamma=4\pi$}
        \label{fig:archimedes_4pi}
    \end{subfigure}
    \caption{Examples of Archimedes' spirals used in this work for the SK mapping scheme. Parameter $\gamma$ controls the robustness of the DNN parameters against channel noise.}
    \label{fig:archimedes_curves}
\end{figure}

Spirals generated by different $\gamma$ parameters are shown in Fig. \ref{fig:archimedes_curves}. Each network parameter $w \in \mathbb{R}$ is mapped to a point $(x_1, x_2)$ on the spiral. Sign of the parameter is encoded by mapping positive-valued DNN parameters to the spiral defined by Eq. (\ref{eq:spiral1}), and the negative-valued ones to the spiral defined by Eq. (\ref{eq:spiral2}). We note that parameter $\gamma$ can effectively control the length of the spirals, and the distance between the negative and positive spirals, which impacts the robustness of this coding technique. At low $\mathrm{SNR}$, a high $\gamma$ value may lead to the two spirals being too close to each other, resulting in sign errors in decoding. However, at high $\mathrm{SNR}$, one can increase $\gamma$ in order to better allocate the 2D space spanned by the spirals.

After the transmission, we decode the original parameters by mapping the received symbols $(y_1, y_2)$ back to the values of network parameters by finding the nearest point on either of the spirals, as shown below:
\begin{equation}
    \hat{w} = \pm \underset{w}{\operatorname{argmin}} \left((y_1 - \theta(w))^2 + (y_2 - \theta(w))^2 \right),
\end{equation}
where $\theta(\cdot)$ represents the union of the spirals defined by Eq. (\ref{eq:spiral1}), and (\ref{eq:spiral2}).

We note that the SK expansion as defined above only allows a $1:2$ bandwidth expansion ratio. In order to achieve higher orders of expansion, one may consider re-applying the same expansion to $x_1$ and $x_2$, by simply replacing $w$ in Eq. (\ref{eq:spiral1}) and Eq. (\ref{eq:spiral2}) by $x_1$ and $x_2$. With this, we can implement a bandwidth ratio of $1:2^n$, where $n$ is the number of expansion steps applied. In order to achieve more flexibility in the overall expansion rates, we propose two methods, which allow to achieve intermediate expansion levels by applying different expansion rates to each layer of the network, depending on the available bandwidth. For a detailed explanation of the methods, please see Section \ref{subs:uneven_repetition}.

An alternative, much simpler method is \textit{layer repetition}. In this method, we simply transmit each network parameter multiple times and average the outputs at the receiver in order to reduce the variance of the noise component. We note that channel repetition can effectively achieve rates of expansion of $1:n$; thus, it is inherently more flexible than SK expansion; however, it does not exploit the higher-dimensional space as effectively as SK expansion, which leads to a sub-optimal performance as we will observe in Section \ref{sec:results}.

\subsection{UEP of DNN parameters against channel noise}
\label{subs:uneven_repetition}

The main limitation of the methods explained above is that they only consider uniform expansion of the entire network. In this section, we aim at providing methods that achieve intermediate levels of network expansion, and better accommodate the available bandwidth. It is known that the impact of different DNN layers on the overall performance varies \cite{hawq}. Therefore, to exploit this inhomogeneity, we will look at methods that apply different expansion ratios to each layer of a DNN. A proper selection of the layers that should be expanded is extremely important in order to achieve satisfactory performance, thus a sensitivity metric is necessary to specify which layers should be protected more than the others.

The first sensitivity metric we propose is based on the variation in the loss function imposed by injecting a certain amount of noise into a DNN, one layer at a time. We hypothesize that the layers that lead to a higher increase in the loss function when perturbed are more sensitive, and hence, should be protected more. Let $\tilde{\mathbf{w}}^{(i)}$ denote the network $\mathbf{w}$ with certain amount of noise injected into its $i$-th layer. Our sensitivity measure is based on the squared difference between the loss function $l(\mathbf{w}, I)$ of the original DNN and the same network when the $i$-th layer is perturbed: $l(\tilde{\mathbf{w}}^{(i)}, I)$. The sensitivity of the $i$-th layer can be expressed as:

\begin{equation}
    s_1^{(i)} = \sum_{I_j \in \mathcal{D}} \left(l(\mathbf{w}, I_j) - l(\tilde{\mathbf{w}}^{(i)}, I_j) \right)^2,
\end{equation}
where $\mathcal{D}$ denotes the training set available at the edge server.

Next, we consider a Hessian-based sensitivity metric \cite{hawq}, where the largest eigenvalue $\lambda_i$ of the Hessian matrix associated with a given layer $i$ is treated as the sensitivity metric of this layer. Since the computation of the Hessian matrix is not feasible for large DNNs, we follow the Von Mises iteration \cite{power_iteration} to estimate it, as shown in Algorithm 1. Once the eigenvalue $\lambda_i$ is calculated for the $i$-th layer, we simply set its sensitivity as $s_2^{(i)} = \lambda_i$.

\begin{algorithm}[t]
\caption{Von Mises iteration for calculating the largest eigenvalue of the Hessian matrix associated with layer $\mathbf{w}_i$.}
\SetAlgoLined
\KwIn{$i$-th layer $\mathbf{w}_i$ of a DNN $\mathbf{w}$, training dataset $\mathcal{D}$.}

Calculate the loss $L = \sum_{I_j \in \mathcal{D}} l(\mathbf{w}, I_j)$;

Calculate the gradient $\mathbf{g}_i = \frac{\partial L}{\partial \mathbf{w}_i}$ of the loss w.r.t. $\mathbf{w}_i$;

Draw a random vector $\mathbf{v}_i$ of the same dimension as $\mathbf{w}_i$; 

Normalize $\mathbf{v}_i$: $\mathbf{v}_i \gets \frac{\mathbf{v}_i}{\norm{\mathbf{v}_i}_2}$.

\Repeat{$\lvert \mathbf{v}_i - \mathbf{v}_i^{(prev)}\rvert < \epsilon$}
{
Calculate inner product $\mathbf{g}_i^T \mathbf{v}_i$;

Calculate the Hessian and $\mathbf{v}_i$ product by $H_i\mathbf{v}_i = \frac{\partial (\mathbf{g}_i^T \mathbf{v}_i)}{\partial \mathbf{w}_i}$;

$\mathbf{v}_i^{(prev)} \gets \mathbf{v}_i$;

Update $\mathbf{v}_i$: $\mathbf{v}_i \gets \frac{H_i\mathbf{v}_i}{\norm{H_i\mathbf{v}_i}_2}$;

Calculate $\lambda_i = \frac{\mathbf{v}_i^T H_i \mathbf{v}_i}{\mathbf{v}_i^T \mathbf{v}_i}$.
}
\KwOut{Largest eigenvalue $\lambda_i$ of the Hessian matrix associated with layer $\mathbf{w}_i$.}

\end{algorithm}

Expanding the layer with the highest sensitivity may not always result in the best performance. This is because some layers contain more parameters than others, and expanding them requires more bandwidth than some less sensitive, yet already compact layers. Similarly, some layers may have higher variance than others, and expanding them may lead to a significant increase in the average power. To better allocate the available power and bandwidth resources, we normalize each sensitivity parameter by the total energy of the corresponding layer, $\tilde{s}_j^{(i)} \triangleq s_j^{(i)}/ \norm{\mathbf{w}_i}_2^2, j=1,2$. Once the sensitivities are calculated for all the layers, we iteratively expand the layers with the highest sensitivity until the available bandwidth is exhausted, as shown in Algorithm 2.

\begin{algorithm}[h]
\caption{Uneven bandwidth expansion based on layer sensitivities.}
\SetAlgoLined
\KwIn{Layers $\mathbf{w}_i \in \mathbb{R}^{d_i}$ of a DNN and their sensitivities $s^{(i)}$, bandwidth $b$.}

Initialize $r_i=1,\ \forall_i$.

\While{$\sum_i r_id_i \leq b$}
{
Calculate normalized sensitivities $\tilde{s}^{(i)} = \frac{s^{(i)}}{r_i\norm{\mathbf{w}_i}_2^2}$;

Find $i^* = \underset{i}{\operatorname{argmax}}\ \tilde{s}^{(i)}$.

$r_i \gets r_i + 1$ (or $r_i \gets 2r_i$ if SK expansion is considered).
}

\KwOut{Number of repetitions $\{r_i\}$ for the selected sensitivity metric.}

\end{algorithm}

\section{SNR Robustness}
\label{sec:snr_robustness}

So far, training has been done targeting a particular channel SNR. With this approach, AirNet performs best if the mismatch between training and test SNRs is minimal. However, such a solution is not practical as it requires the edge server to store multiple sets of DNN parameters, each trained for a specific SNR value. In this section, we will present two methods to significantly reduce memory requirements.

In the first scheme, we train a single set of DNN parameters to be used over a range of SNR values from the interval $\left[\textrm{SNR}_{min}, \textrm{SNR}_{max}\right]$. Instead of sampling a noise vector from a fixed SNR target, we consider a different noise variance at every training iteration. SNR values over the iterations are chosen according to the ``sandwich rule'', which is frequently used in the efficient DNN design literature \cite{slimmable_nets}. In the first training iteration, we target $\textrm{SNR}_{min}$, in the second iteration $\textrm{SNR}_{max}$, and finally, in the third iteration, we target an SNR value randomly sampled from $\left[\textrm{SNR}_{min}, \textrm{SNR}_{max}\right]$, i.e., $\textrm{SNR}=\textrm{SNR}_{interp}\sim \mathcal{U}\left(\textrm{SNR}_{min}, \textrm{SNR}_{max}\right)$. We repeat these three iterations throughout the entire training process to make sure the final DNN can adapt to a variety of SNR values it can experience during testing.

\begin{figure}[t]
\begin{center}
\includegraphics[width=.9\linewidth]{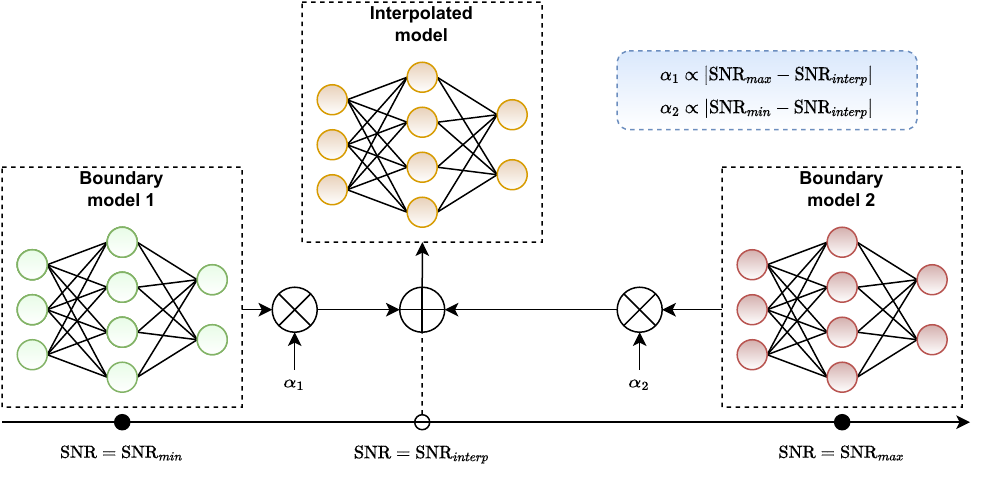}
\end{center}
  \caption{Proposed interpolation scheme. Given a channel SNR, the interpolated network is obtained as a weighted sum of the boundary networks with weights $\alpha_1$ and $\alpha_2$.}
\label{fig:interpolation_scheme}
\end{figure}

Despite its simplicity, the above method is limited, as it still relies on a single set of parameters, which cannot perfectly adapt to every SNR. To overcome this limitation, we consider an ensemble training approach inspired by \cite{loss_surfaces}, which stores only two sets of DNN parameters at the edge server. In this \textit{interpolation scheme} (see Fig. \ref{fig:interpolation_scheme}), the two sets of DNN parameters, $\mathbf{w}_{min}$ and $\mathbf{w}_{max}$, called the \textit{boundary networks}, are trained targeting channel SNR values $\textrm{SNR}_{min}$ and $\textrm{SNR}_{max}$. In the first iteration, we set the SNR target to $\textrm{SNR}_{min}$, and train only the boundary network $\mathbf{w}_{min}$. In the second iteration, we repeat this process for the boundary network $\mathbf{w}_{max}$ by setting $\textrm{SNR}=\textrm{SNR}_{max}$. Finally, in the third training iteration, we consider a random SNR value from $\left[\textrm{SNR}_{min}, \textrm{SNR}_{max}\right]$, i.e., $\textrm{SNR}=\textrm{SNR}_{interp}\sim \mathcal{U}\left(\textrm{SNR}_{min}, \textrm{SNR}_{max}\right)$, and train a DNN with parameters $\mathbf{w}_{interp}$ equal to the weighted sum of the boundary networks' parameters. The exact values of the interpolated model parameters are calculated as:

\begin{equation}
\label{eq:interpolation_weighting}
    \mathbf{w}_{interp} = \alpha_1 \mathbf{w}_{min} + \alpha_2 \mathbf{w}_{max},
\end{equation}
where $\alpha_1 = \frac{|\textrm{SNR}_{max} - \textrm{SNR}_{interp}|}{|\textrm{SNR}_{max} - \textrm{SNR}_{min}|}$ and we set $\alpha_2 = 1 - \alpha_1$.

This strategy allows us to train a family of interpolated networks that achieve satisfactory performance on a range of SNR values between $\textrm{SNR}_{min}$ and $\textrm{SNR}_{max}$. In this work, we initialize both boundary networks with the same set of weights, and they naturally converge to different optima as their parameters are updated with different SNRs. Once the networks are trained, during the test phase, depending on the experienced channel SNR we can sample an interpolated network, and also apply SK mapping or repetition schemes to it as desired, in order to further increase the performance.

\section{Results}
\label{sec:results}
\subsection{Experimental setup}
\label{subs:experimental_setup}

In this work, we focus on transmitting parameters of DNNs trained for the image classification task. For evaluation, we utilize two distinct datasets. The first dataset we consider is CIFAR10 \cite{cifar}, which consists of 60000 RGB images of $32\times 32$ resolution. The images represent 10 different classes. Following the standard protocol, we utilize 50000 images for training and 10000 for testing, and employ the top-1 classification accuracy as our primary accuracy metric. For a fair comparison with other approaches, we consider Small-VGG16 \cite{vgg} as our baseline DNN. The Small-VGG16 follows a similar structure to the standard VGG16 network, but replaces the standard classifier head with a new one that consists of three fully-connected layers, the first two containing 512 neurons with ReLU activation, and the third containing 10 output neurons for class prediction. During training we use cross-entropy loss, stochastic gradient descent (SGD) optimizer with a learning rate of $0.01$ and momentum of $0.9$ for 30 epochs, reduce the learning rate to $0.001$ and train for a further 30 epochs. The same procedure applies to both initial training and fine-tuning after every pruning step.

The second dataset we utilize in this work is Tiny ImageNet \cite{tiny_imagenet}. The dataset consists of 100000 $64\times 64$ training images of 200 classes, where each class is uniformly represented with 500 image samples. The test set consists of 10000 images, and, as before, we use top-1 classification accuracy as our performance metric. We adopt ResNet-34 \cite{resnet} as our DNN architecture for this task. During training, we utilize cross-entropy loss, and SGD optimizer with a learning rate of $0.001$ and momentum of $0.9$, reduce the learning rate to $0.0001$ after 60 epochs and train for a further 30 epochs. In order to sustain the fixed channel depth after every block of ResNet, during the pruning phase, we only prune the output channels of the first convolutional layer of each ResNet block. Before feeding Tiny ImageNet images into ResNet, we resize them to the resolution of $256\times 256$ and crop the central $224\times 224$ pixels from each image.

We perform multiple training runs of the networks for different $\textrm{SNR}$ values, available bandwidth constraints $b$, channel models, and training strategies. Unless specified differently, in the fading channel scenario, we assume the CSI is available only at the receiver.

We compare our scheme against two separation-based schemes: DeepCABAC \cite{deepCABAC} and SuRP \cite{succesive_pruning}. Both methods first perform network pruning to obtain a sparse structure, which is followed by network quantization and compression via either arithmetic or Huffman coding. For both methods, we performed multiple runs of pruning and compression with various hyperparameters, and considered only the best-performing ones in our evaluations. When digital schemes are used over the fading channel, we consider two scenarios regarding CSI availability. In the first scenario, we consider the CSI is only available at the receiver, and the transmitter is assumed to transmit at a fixed rate. If the channel capacity is below this rate, we assume the transmission fails, i.e., an outage occurs. We then calculate the fraction of successful transmissions and multiply it by the accuracy achieved by the transmitted DNN. In the second scenario, we consider the availability of CSI at both the transmitter and the receiver. This will be denoted by CSIT in the simulations. In this scenario, the transmitter always transmits at the capacity of the channel, which serves as an upper-bound on the performance of separation-based DNN delivery schemes.

\subsection{Performance comparison}

In this section, we present the comparison between AirNet and alternative separation-based schemes. For AirNet, we consider three alternatives - vanilla AirNet, denoted simply as AirNet, which employs noise injection, pruning, and KD, AirNet with SK bandwidth expansion and UEP (denoted as AirNet + SK + UEP) presented in Section \ref{subs:uneven_repetition} with Hessian-based sensitivity, and AirNet trained with the interpolation scheme (presented in Section \ref{sec:snr_robustness}), with SK scheme and uneven error protection (denoted as AirNet + I + SK + UEP).

\begin{figure*}[t]
    \centering
    \begin{subfigure}[t]{0.49\textwidth}
        \centering
        \includegraphics[width=\textwidth]{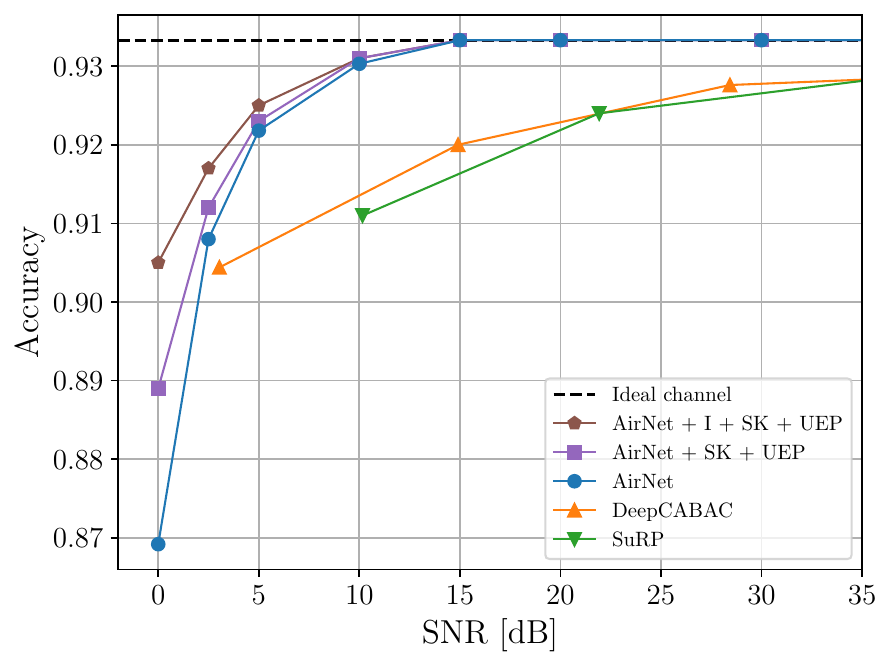}
        \caption{AWGN, $b\approx1.2\times 10^6$}
        \label{fig:awgn_comparison}
    \end{subfigure}
    \begin{subfigure}[t]{0.49\textwidth}
        \centering
        \includegraphics[width=\textwidth]{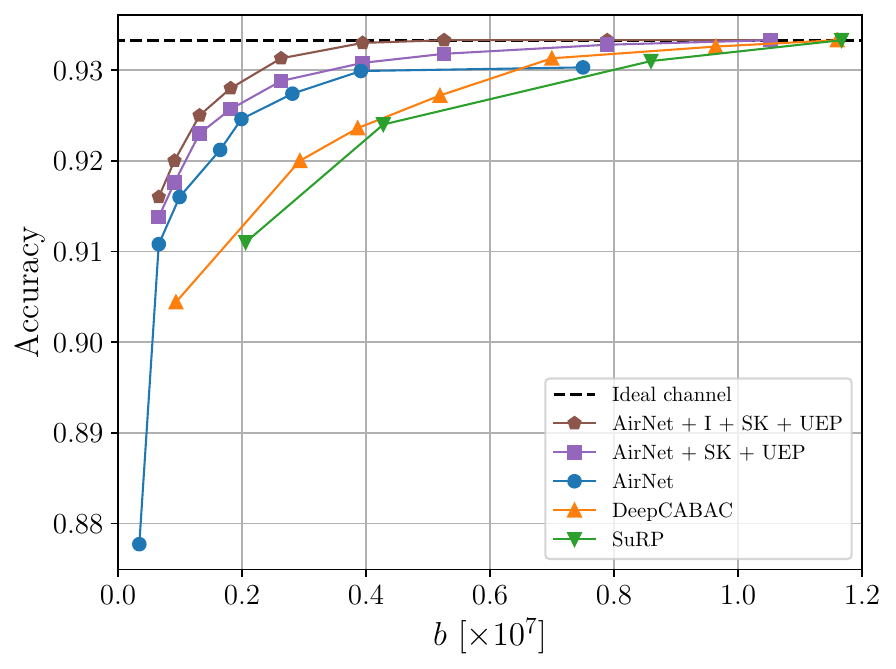}
        \caption{AWGN, $\text{SNR}=5\textrm{dB}$}
        \label{fig:awgn_bandwidth}
    \end{subfigure}
    \begin{subfigure}[b]{0.49\textwidth}
        \centering
        \includegraphics[width=\textwidth]{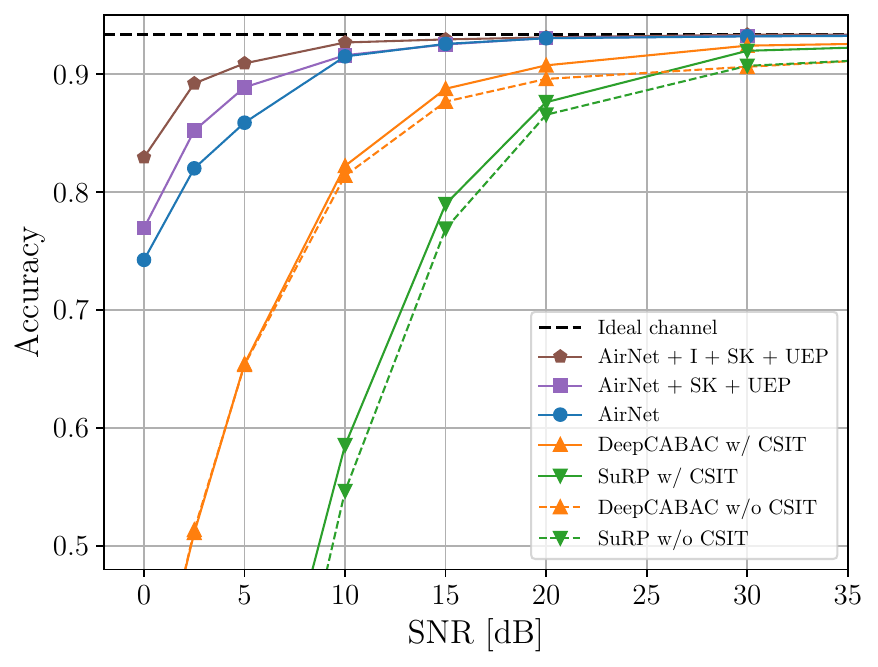}
        \caption{Fading, $b\approx1.2\times 10^6$}
        \label{fig:fading_comparison}
    \end{subfigure}
    \begin{subfigure}[b]{0.49\textwidth}
        \centering
        \includegraphics[width=\textwidth]{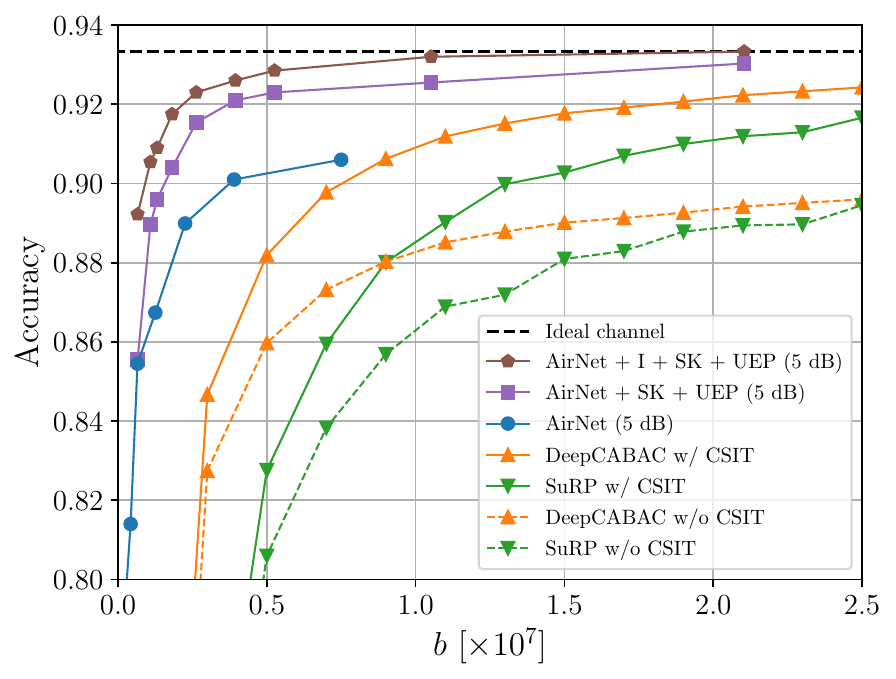}
        \caption{Fading, $\text{SNR}=5\textrm{dB}$}
        \label{fig:fading_bandwidth}
    \end{subfigure}
    \caption{Performance comparison between the proposed AirNet approaches, and the alternative digital and analog schemes over AWGN and slow fading channels for a range of channel SNRs and bandwidths in image classification task with Small-VGG16 network and CIFAR10 dataset. 
    }
    \label{fig:methods_comparison}
\end{figure*}

\textbf{Small-VGG16.} In Fig. \ref{fig:methods_comparison}, we consider the image classification task with the Small-VGG16 network. In Fig. \ref{fig:awgn_comparison} we fix the bandwidth $b$ to approximately $1.2\times 10^6$ channel uses and vary the SNR of an AWGN channel. For the AirNet and AirNet + SK + UEP curves, each point is achieved by a separate model trained specifically for the corresponding SNR value used for testing. On the other hand, for AirNet + I + SK + UEP, each point corresponds to a DNN obtained as the weighted sum of two boundary models, as explained in Section \ref{sec:snr_robustness}. For the separation-based DeepCABAC and SuRP approaches, the DNN is compressed to the level allowed by channel capacity. We see that AirNet is able to outperform separation-based alternatives for every value of the SNR by a large margin. This is despite the fact that we assumed capacity-achieving channel coding for the separation-based approaches. More strikingly, AirNet is able to achieve satisfactory accuracy even at extremely low values of SNR. This accuracy is further improved by the use of SK mapping with the UEP scheme. This is particularly beneficial in the low SNR regime. The proposed interpolation scheme not only removes the requirement of training a separate model for every target SNR, but also brings further performance improvement, especially at low $\mathrm{SNR}$ values. Our scheme is able to recover the original accuracy of the network at a moderately low SNR value of $15\mathrm{dB}$, whereas the digital alternatives achieve significantly lower accuracy even at the SNR of $35\mathrm{dB}$.

Similar trends can be observed when we fix the SNR value to $5\mathrm{dB}$ and vary the available bandwidth $b$. AirNet is able to outperform digital alternatives at every bandwidth value considered. We note, that SK bandwidth expansion improves the network performance, which indicates that it is better to first prune the network below the available bandwidth, and to further expand it for better protection of the remaining DNN parameters. The vanilla AirNet without SK bandwidth expansion is not able to recover the original DNN accuracy, whereas, with bandwidth expansion, which can exploit the available channel bandwidth for better protection against noise, the original accuracy of the DNN can be recovered at $b\approx 8 \times 10^6$. The digital alternatives are consistently outperformed by the proposed schemes, as they require much larger bandwidth to achieve similar levels of accuracy.

\begin{figure*}[ht]
    \centering
    \begin{subfigure}[t]{0.49\textwidth}
        \centering
        \includegraphics[width=\textwidth]{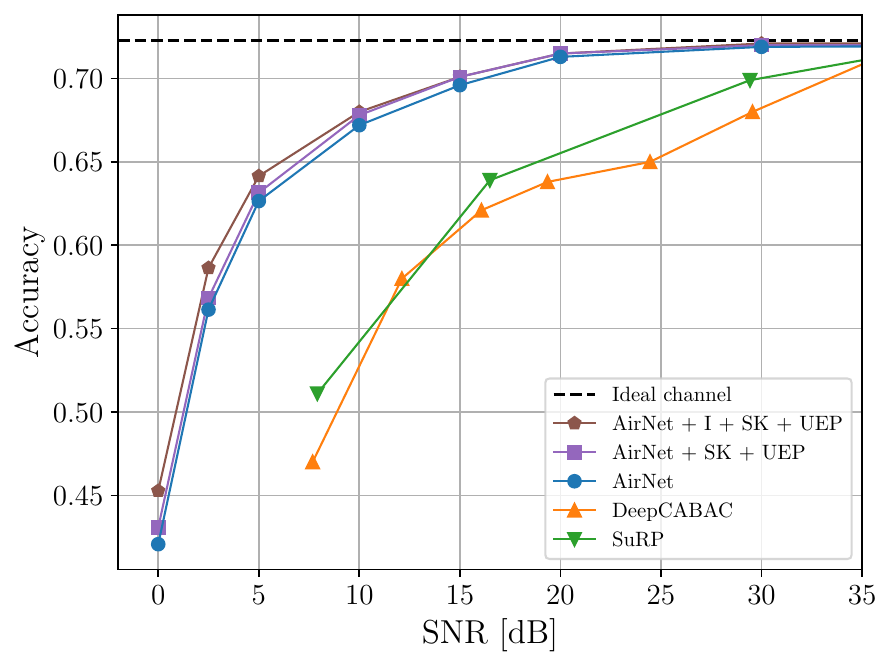}
        \caption{AWGN, $b\approx4.5\times 10^6$}
        \label{fig:awgn_comparison_resnet}
    \end{subfigure}
    \begin{subfigure}[t]{0.49\textwidth}
        \centering
        \includegraphics[width=\textwidth]{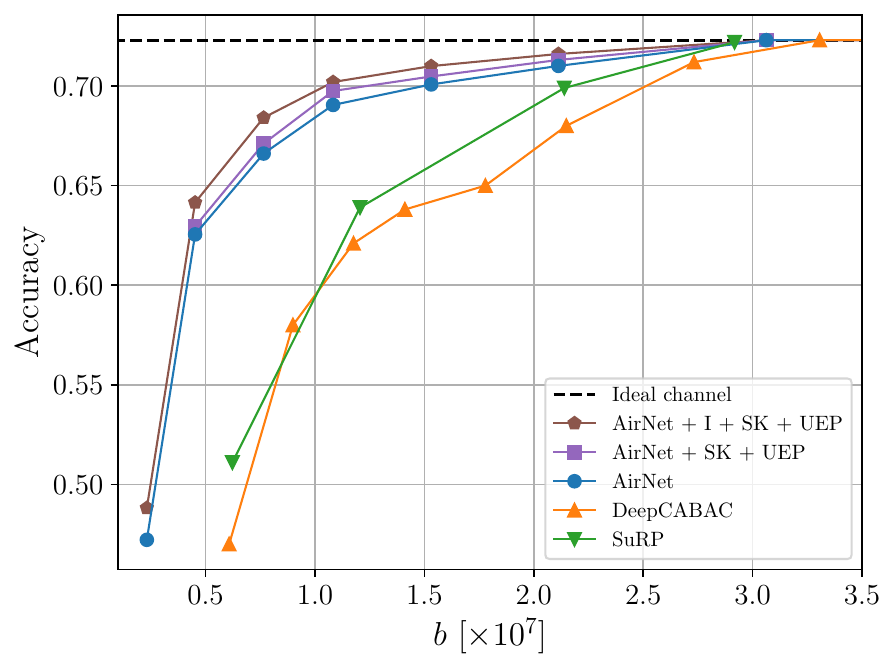}
        \caption{AWGN, $\text{SNR}=5\textrm{dB}$}
        \label{fig:awgn_bandwidth_resnet}
    \end{subfigure}
    \begin{subfigure}[b]{0.49\textwidth}
        \centering
        \includegraphics[width=\textwidth]{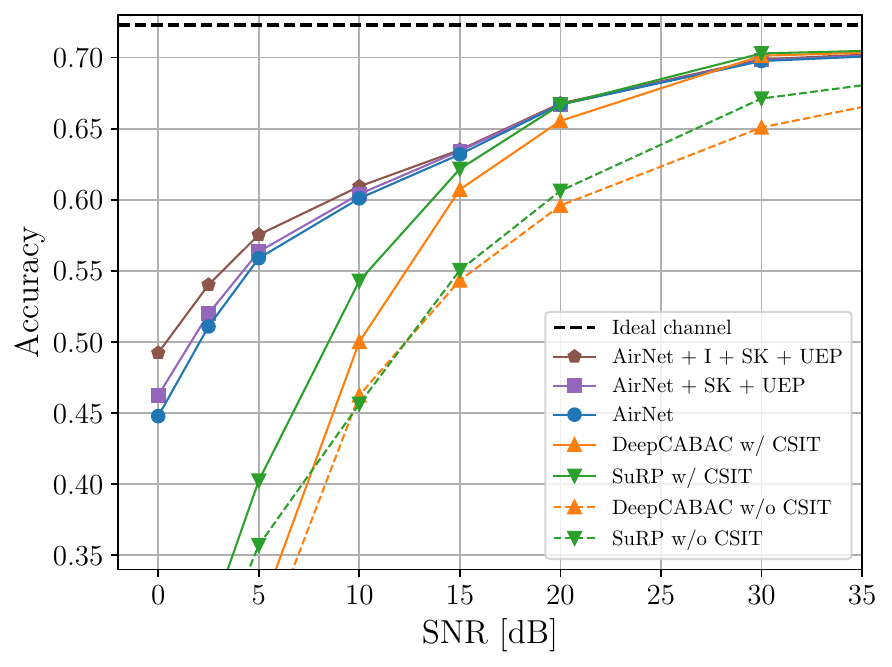}
        \caption{Fading, $b\approx7.5\times 10^6$}
        \label{fig:fading_comparison_resnet}
    \end{subfigure}
    \begin{subfigure}[b]{0.49\textwidth}
        \centering
        \includegraphics[width=\textwidth]{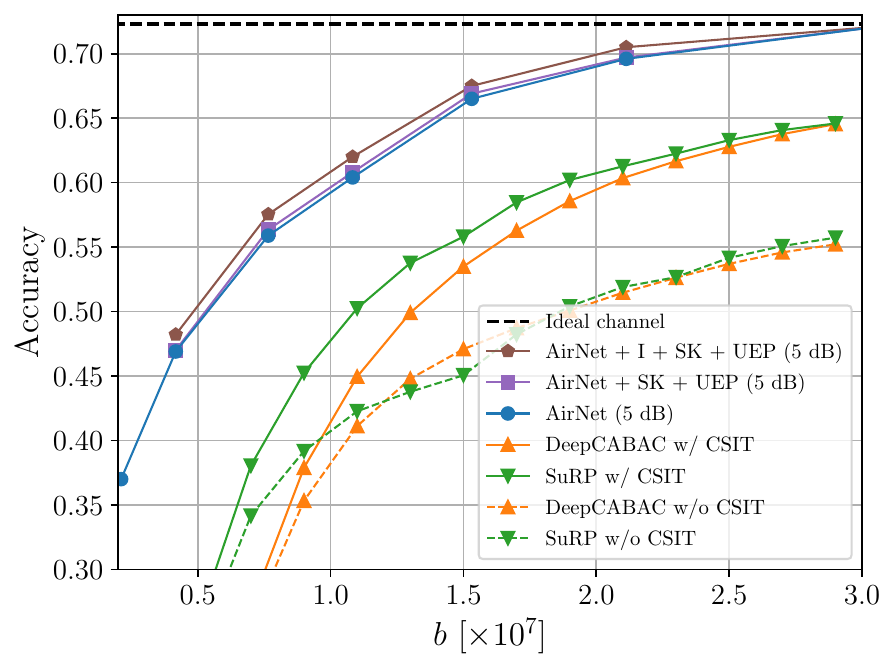}
        \caption{Fading, $\text{SNR}=5\textrm{dB}$}
        \label{fig:fading_bandwidth_resnet}
    \end{subfigure}
    \caption{Performance comparison between the proposed AirNet approaches, and the alternative digital and analog schemes over AWGN and slow fading channels for a range of channel SNRs and bandwidths in classification task with ResNet-34 network and Tiny ImageNet dataset. Our training strategy generalizes well to different network architectures and datasets.}
    \label{fig:methods_comparison_resnet}
\end{figure*}

We consider fading channels with different average SNR values, and a fixed bandwidth of $b=1.2\times 10^6$ in Fig. \ref{fig:fading_comparison}. Here, each point of the AirNet curves is obtained by training a different DNN to be used over the fading channels with that particular average SNR value. Again, AirNet is able to achieve better accuracy than the separation-based methods, while also being able to recover the original DNN accuracy at an average SNR of $20\mathrm{dB}$. The improvement of AirNet is even more evident against digital schemes without CSI. We remind that the AirNet scheme does not assume CSI at the transmitter. These schemes experience sharp accuracy drop whenever the SNR drops below $\sim 15\mathrm{dB}$. As already observed in the AWGN case, SK bandwidth expansion helps to increase the robustness of the network, especially at the low values of SNR, which can be further improved by the interpolation scheme, which also enjoys low memory requirements.

In Fig. \ref{fig:fading_bandwidth}, we fix the average SNR value to $5\mathrm{dB}$ in the fading regime, while varying the channel bandwidth $b$. We observe that both vanilla AirNet and AirNet with SK bandwidth expansion are able to achieve satisfactory accuracy at a wide range of channel bandwidths. Separation-based schemes tend to outperform AirNet without SK expansion when they have access to CSIT. The results here further motivate the use of bandwidth expansion as its benefit is clear over fading channels in the low SNR regime. It can be observed that at $\mathrm{SNR}$ of $5\mathrm{dB}$, the interpolation scheme still yields performance improvements over a wide range of bandwidths tested. 

\textbf{ResNet-34.} In Fig. \ref{fig:methods_comparison_resnet}, we present the results for the classification task on the Tiny ImageNet dataset with ResNet-34 network. ResNet-34 architecture differs significantly from Small-VGG16 as it is a much deeper architecture with residual connections and batch normalization layers.

In Fig. \ref{fig:awgn_comparison_resnet}, we first fix the available bandwidth $b$ to approximately $4.5\times10^6$ channel uses and train a separate network for a variety of different SNR values. Our observations are consistent with the previous experiments with Small-VGG16. Networks trained with AirNet outperform separation-based counterparts by a large margin, being able to recover the original network accuracy at $\mathrm{SNR}$ of $30\mathrm{dB}$ in a significantly more challenging classification task. SK mapping combined with UEP scheme is able to bring improvements over the vanilla AirNet, although the gains are more limited in this scenario. Similarly with the previous experiments, the interpolation scheme increases the accuracy within low $\mathrm{SNR}$ regime, compared to SK + UEP, while still significantly reducing storage requirements.

Results presented in Fig. \ref{fig:awgn_bandwidth_resnet} show that our strategy consistently outperforms digital alternatives at a wide range of channel bandwidths $b$ for a fixed $\mathrm{SNR}$ of $5\mathrm{dB}$. Small improvements are achieved through applying SK expansion with the UEP scheme, which can be pushed further by applying the interpolation scheme.

Our method shows good generalizability to fading channel scenarios as well, and significantly outperforms separation-based alternatives, both for fixed bandwidth $b$ (see Fig. \ref{fig:fading_comparison_resnet}), and fixed SNR (see Fig. \ref{fig:fading_bandwidth_resnet}) scenarios. It is further observed that the SK expansion with UEP leads to performance improvements as before, which can be amplified by the use of the interpolation scheme, especially at low SNRs. For $\mathrm{SNR}$ values above $5\mathrm{dB}$, no significant difference is observed between different variations of AirNet, meaning that the proposed extensions are effective mainly in the low SNR regime.

\begin{figure*}[t]
    \centering
    \begin{subfigure}[t]{0.49\textwidth}
        \centering
        \includegraphics[width=\textwidth]{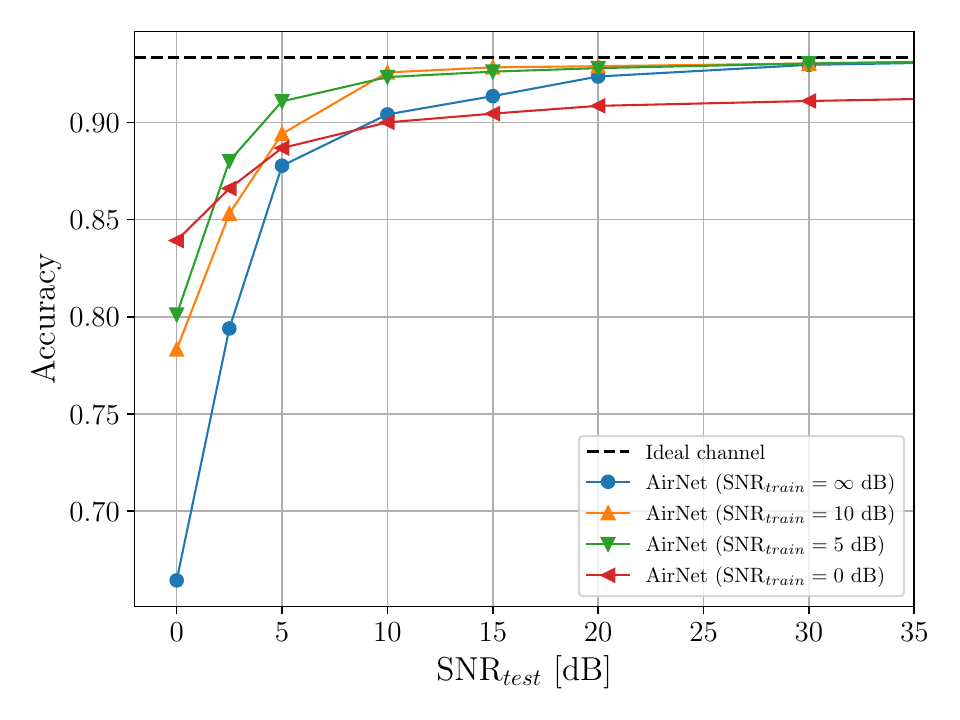}
        \caption{Small-VGG16, CIFAR10, $b\approx0.65\times10^6$}
        \label{fig:snr_mismatch_classification}
    \end{subfigure}
    \begin{subfigure}[t]{0.49\textwidth}
        \centering
        \includegraphics[width=\textwidth]{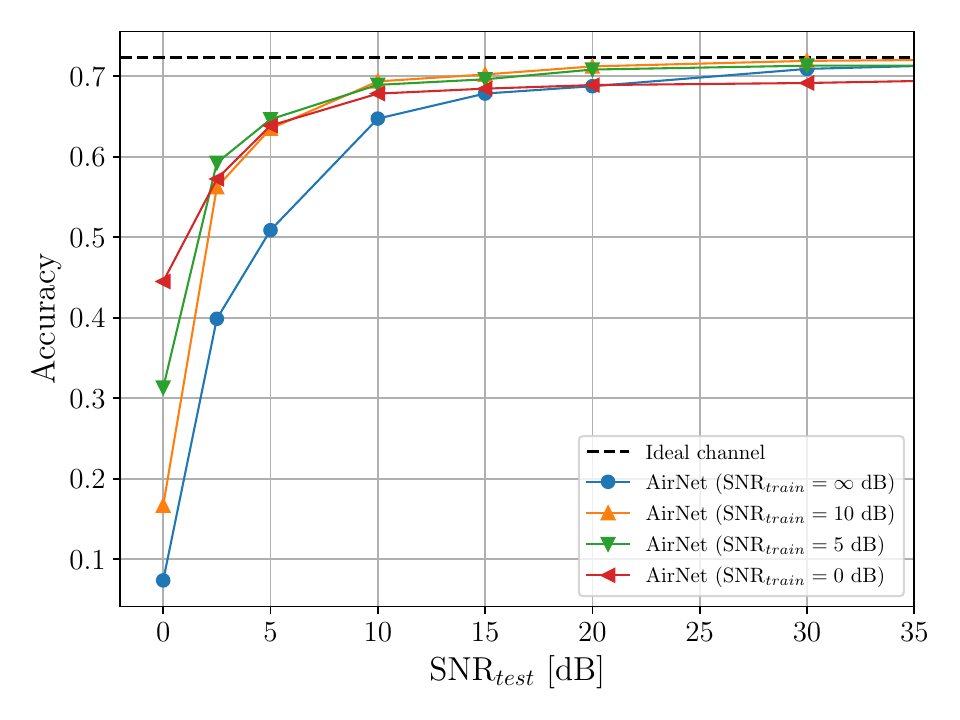}
        \caption{ResNet-34, Tiny ImageNet, $b\approx6.5\times10^6$}
        \label{fig:snr_mismatch_resnet}
    \end{subfigure}
    \caption{Performance comparison of the models trained for different values of $\mathrm{SNR}_{train}$ and tested on a range of $\mathrm{SNR}_{test}$.
    }
    \label{fig:snr_mismatch}
\end{figure*}

\subsection{Generalizability to different noise SNR}

So far, we have assumed that the DNN trained for a specific target SNR is tested at that SNR. In this section, we investigate the effects of the mismatch between the training and test SNRs. In Fig. \ref{fig:snr_mismatch_classification}, we show the performance of the networks trained for different $\mathrm{SNR}_{train}$ values varying between $0\mathrm{dB}$ and $10\mathrm{dB}$ as well as a model trained without noise, denoted as $\mathrm{SNR}_{train}=\infty\ \mathrm{dB}$, when tested with $\mathrm{SNR}_{test}$ values between $0\mathrm{dB}$ and $35\mathrm{dB}$. We see that the model trained and tested at the same SNR, i.e., $\mathrm{SNR}=\mathrm{SNR}_{train}=\mathrm{SNR}_{test}$, achieves the best performance. However, models trained with moderate values of $\mathrm{SNR}_{train}$ seem to achieve reasonable performance for a wide range of different $\mathrm{SNR}_{test}$. The model trained without noise injection ($\mathrm{SNR}_{train}=\infty\ \mathrm{dB}$) clearly suffers in the low $\mathrm{SNR}_{test}$ regime. This indicates the necessity of the noise injection step throughout training. We note that standard separation-based transmission schemes usually exhibit \textit{cliff effect}, i.e., the accuracy sharply degrades when the $\mathrm{SNR}_{test}$ falls below the target SNR value. As observed in Fig. \ref{fig:snr_mismatch_classification}, our model, on the contrary, exhibits \textit{graceful degradation}, thus its accuracy slowly degrades as the channel noise variance increases, which makes AirNet even more desirable in practical implementations, as it is able to achieve good performance even when accurate channel estimation is not possible. In Fig. \ref{fig:snr_mismatch_resnet} we show a similar comparison, but with the ResNet-34 baseline network and Tiny ImageNet dataset. We see that the main observations remain consistent between different architectures and datasets.

\begin{figure}[t]
\begin{center}
\includegraphics[width=0.49\linewidth]{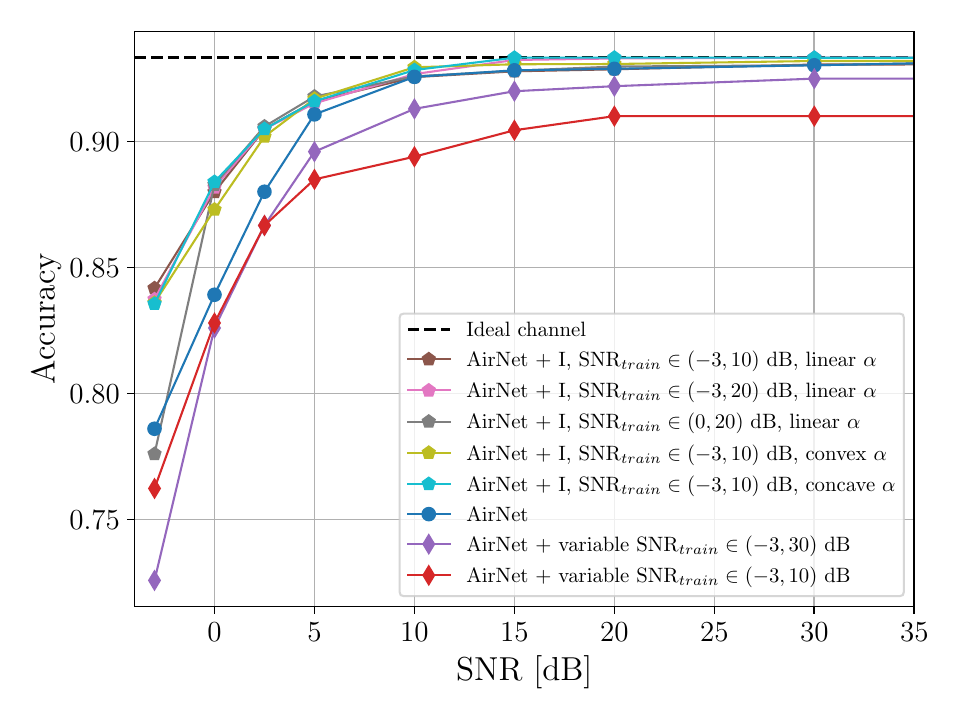}
\end{center}
  \caption{Comparison between vanilla AirNet trained with $\mathrm{SNR}_{train}=\mathrm{SNR}_{test}$, interpolation scheme, and AirNet trained with variable $\mathrm{SNR}$. $b\approx 0.65\times 10^{6}$.
  }
\label{fig:interpolation_vs_variable}
\end{figure}

The above results beg the question of how to train the network when we do not know the channel SNR in advance. The natural approach is to follow the approach used for fading channels, and train the network over a range of SNR values.
In Fig. \ref{fig:interpolation_vs_variable}, we present the accuracy comparison between the three methods as we increase SNR for a fixed bandwidth $b=0.65\times10^6$ channel uses. The blue curve in the figure is the baseline AirNet performance obtained by networks trained for each specific SNR value. One can see that AirNet trained with variable values of SNR (see Section \ref{sec:snr_robustness}) generalizes well to a wide range of SNR, but fails to match the performance of separate networks trained for a specific SNR. Moreover, when a very wide range of $\mathrm{SNR}_{train} \in (-3, 30)\ \mathrm{dB}$ is considered, a significant performance degradation is observed for extremely low values of SNR. Training over a more narrow range of SNR values, $\mathrm{SNR}_{train} \in (-3, 10)\ \mathrm{dB}$, tends to work better in the low SNR regime, yet saturates to a sub-optimal accuracy value as the SNR increases. We observe that the proposed interpolation scheme not only matches the accuracies achieved by separate networks, but introduces further performance improvements at all SNRs, particularly significant in the low SNR regime. This performance improvement does not depend on a specific choice of the $\mathrm{SNR}_{train}$ range, as long as it overlaps with the range of $\mathrm{SNR}_{test}$ values experienced during test, or a shape of the function used to calculate the weighting parameter $\alpha$. In Eq. (\ref{eq:interpolation_weighting}), we proposed $\alpha(\textrm{SNR})$ to be a linear function; however, in Fig. \ref{fig:interpolation_vs_variable}, one can clearly see that similar results can be obtained when the function is replaced with a simple strictly increasing convex or concave function parameterized by a B\'ezier curve. Another important observation is that all the schemes should be trained over an SNR range that includes the SNR values expected to be encountered during transmission. Interpolation scheme suffers from sharp performance degradation when trained at $\mathrm{SNR}_{train} \in (0, 20)\ \mathrm{dB}$ and tested with $\mathrm{SNR}_{test} = -3 \mathrm{dB}$. Furthermore, it is worth noting that, on top of the performance improvements, the interpolation scheme allows for a significant reduction in the number of parameters stored at the edge server, as it only requires storing two sets of DNN parameters for each bandwidth, as opposed to storing a separate set of DNN parameters for each target SNR value at each bandwidth.

\begin{figure*}[t]
    \centering
    \begin{subfigure}[t]{0.49\textwidth}
        \centering
        \includegraphics[width=\textwidth]{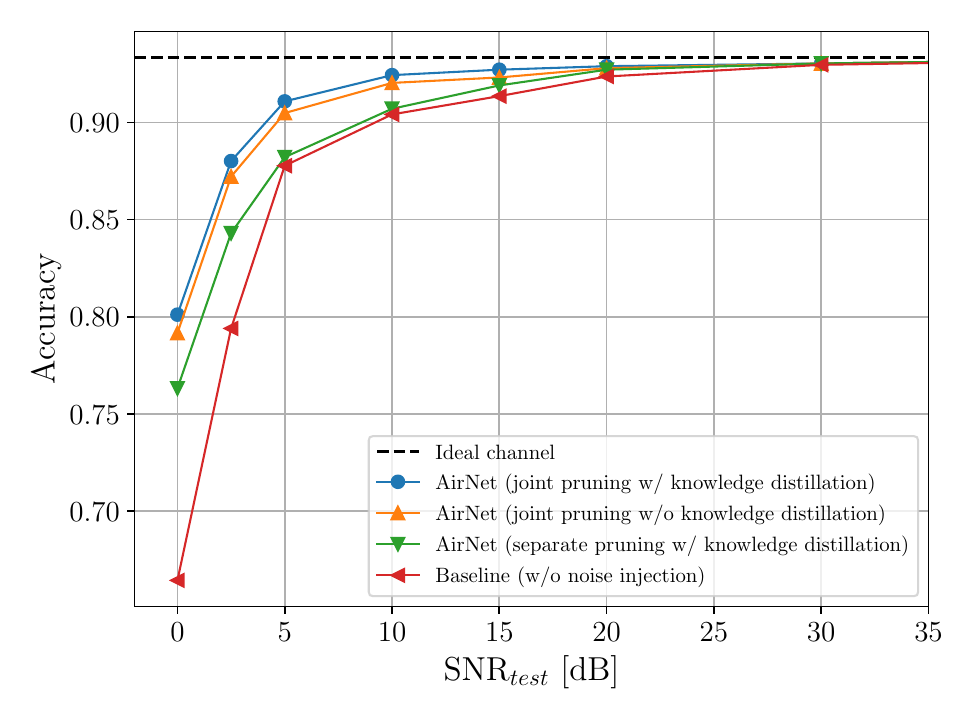}
        \caption{Training strategy, $b\approx 0.65\times 10^{6}$}
        \label{fig:training_steps}
    \end{subfigure}
    \begin{subfigure}[t]{0.49\textwidth}
        \centering
        \includegraphics[width=\textwidth]{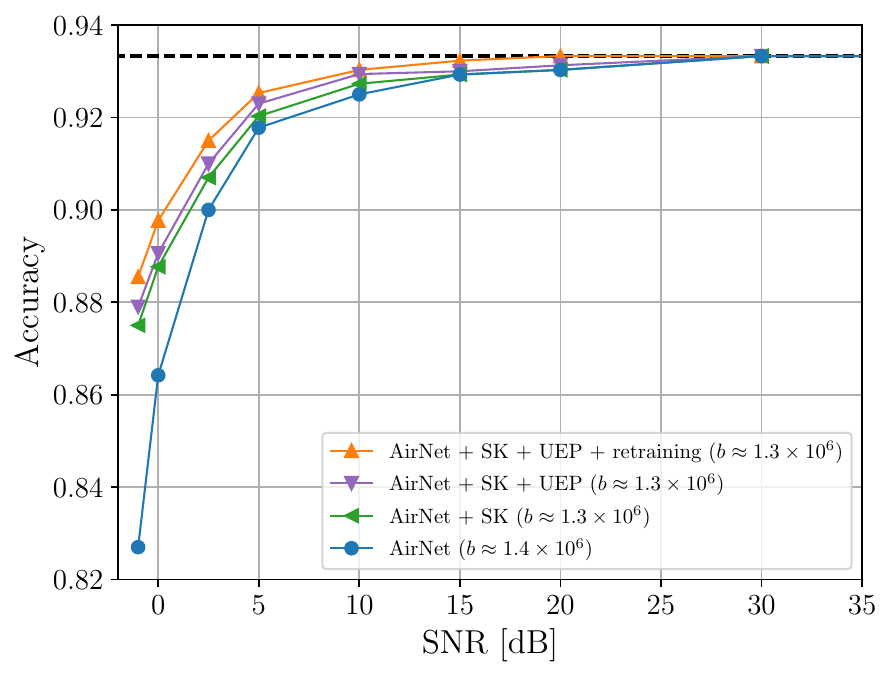}
        \caption{Bandwidth expansion}
        \label{fig:uneven_repetition}
    \end{subfigure}
    \caption{Ablation study for our training strategy (a) and various bandwidth expansion strategies (b). Removing any of the steps from the training strategy leads to a drop in the accuracy across all SNRs. We have $\mathrm{SNR}_{train}=5\mathrm{dB}$.}
\end{figure*}



\subsection{Evaluation of training steps}

In this section, we evaluate the impact of every training step utilized in the AirNet method. A comparison between networks trained with different training strategies is shown in Fig. \ref{fig:training_steps}. One can see that each step is crucial to achieve the best possible final accuracy. The best-performing network utilizes KD and \textit{joint pruning}, meaning that the noise injection is performed jointly with fine-tuning after each pruning iteration. Lack of KD results in a slight decrease in the performance, visible especially at $\mathrm{SNR}_{test}<15\mathrm{dB}$. When we separate pruning from noise injection by first performing pruning, and re-train the network with noise injection afterward, denoted as \textit{separate pruning}, we observe a large drop in the performance for every $\mathrm{SNR}_{test}$ value below $20\mathrm{dB}$. This illustrates that, when a certain amount of noise is injected into the DNN parameters, it effectively learns to prioritize only a few most significant convolutional filters within each layer, whereas the remaining ones get discarded during pruning. Once the pruning process is performed without noise injection, it is impossible for the network to learn this amount of robustness against channel noise. Finally, the model trained without noise injection performs the worst across the entire range.
\begin{figure}[t]
\begin{center}
\includegraphics[width=0.49\linewidth]{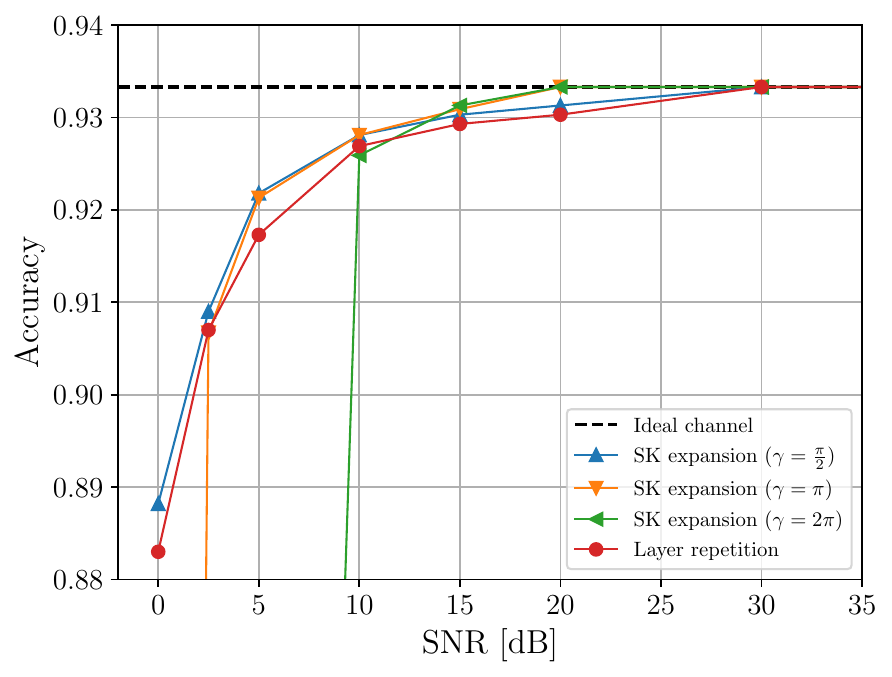}
\end{center}
  \caption{Comparison between SK mapping and layer repetition scheme for bandwidth expansion 
  ($\mathrm{SNR}_{train}=5\mathrm{dB}$, $b\approx 1.3\times 10^{6}$).
  }
\label{fig:sk_vs_repetition}
\end{figure}

\subsection{Comparison of bandwidth expansion methods}


In this section, we compare the different bandwidth expansion methods described in detail in Section \ref{sec:uep}. In Fig. \ref{fig:uneven_repetition}, we show the impact of each bandwidth expansion step on the performance of AirNet. For this comparison, we use vanilla AirNet with noise injection at $\mathrm{SNR}_{train}=5\mathrm{dB}$ and $b\approx1.4\times10^6$. For bandwidth expansion, however, we first prune a DNN to $b\approx0.65\times10^6$ with the same $\mathrm{SNR}_{train}$ and then expand it to $b\approx1.3\times10^6$. We see that the best accuracy is achieved when we perform Hessian-based UEP with the SK bandwidth expansion scheme, which is followed by DNN retraining during which the variance of noise injected into each layer is scaled according to the number of repetitions calculated. We see that the network without retraining (AirNet + SK + UEP) is able to achieve satisfactory accuracy; however, retraining brings further improvement of up to $0.4\%$ accuracy. AirNet without UEP achieves slightly worse performance, but is still superior to the vanilla AirNet, which uniformly protects all the weights, especially at low SNR values.

\begin{figure}[t]
\begin{center}
\includegraphics[width=0.49\linewidth]{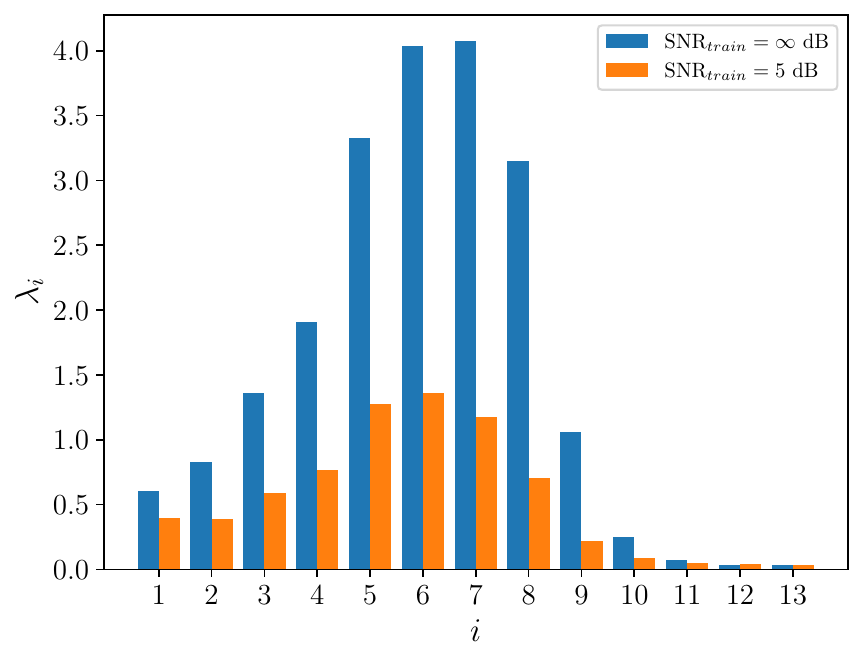}
\end{center}
  \caption{Comparison of per-layer sensitivity measured by the largest eigenvalue $\lambda_i$ of the Hessian matrix corresponding to layer $i,\ i=1,\dots,13$. Training with noise injection significantly reduces the sensitivity of DNN layers ($\mathrm{SNR}_{train}=5\mathrm{dB}$, $b\approx 0.65\times 10^{6}$).
  }
\label{fig:sensitivity}
\end{figure}

In Fig. \ref{fig:sk_vs_repetition}, we consider the SK bandwidth expansion scheme with different values of the $\gamma$ parameter, and the simple layer repetition scheme. We see that even for a relatively low value of $\gamma$, SK expansion outperforms layer repetition for a wide range of SNR values. The performance of the SK expansion scheme can be further improved in the high SNR regime by increasing the $\gamma$ value. However, this results in a sharp accuracy drop in the low SNR regime. This is caused by the fact that large $\gamma$ causes the distance between positive and negative spirals to be small, which results in large positive DNN parameters being decoded as negative values, and vice versa. This means that $\gamma$ is another hyperparameter of the proposed AirNet scheme with SK bandwidth expansion, which can benefit from the availability of CSI at the transmitter.

\subsection{Comparison of different sensitivity estimation criteria}

In this section, we analyze the different sensitivity criteria used in adaptive layer expansion as introduced in Section \ref{subs:uneven_repetition}. In Fig. \ref{fig:sensitivity}, we compare the Hessian-based sensitivity for a network, when trained with and without noise injection. We observe that training with noise injection significantly reduces the sensitivity of the layers, which is extremely beneficial for further wireless transmission of the network.

\begin{figure*}[t]
    \centering
    \begin{subfigure}[t]{0.49\textwidth}
        \centering
        \includegraphics[width=\textwidth]{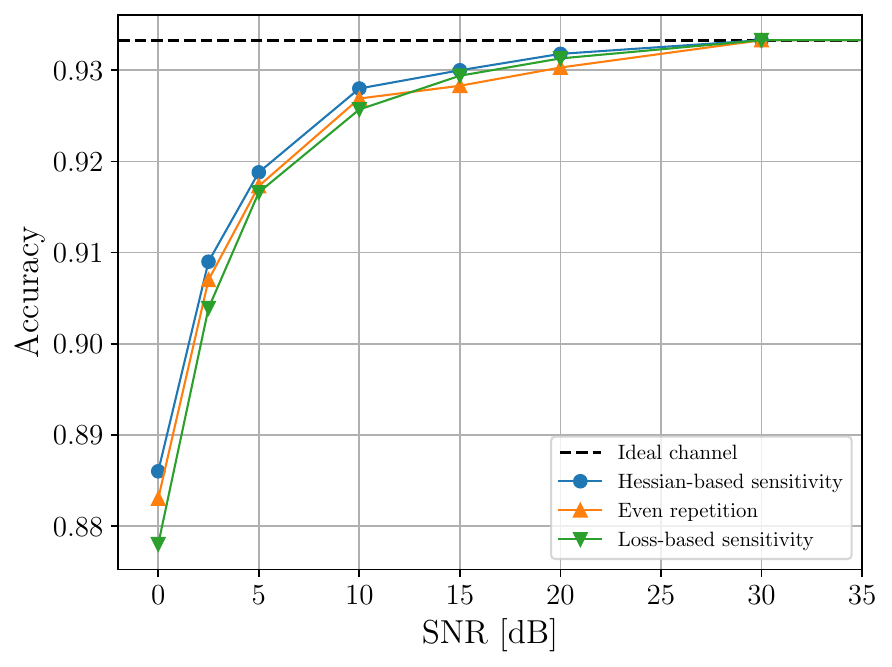}
        \caption{Performance comparison}
        \label{fig:uneven_accuracy}
    \end{subfigure}
    \begin{subfigure}[t]{0.49\textwidth}
        \centering
        \includegraphics[width=\textwidth]{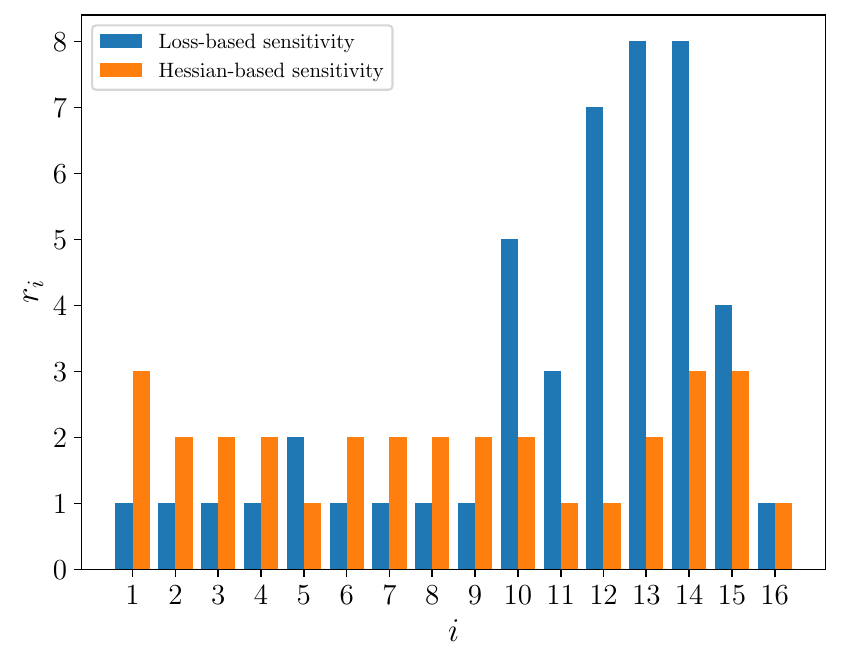}
        \caption{Number of repetitions}
        \label{fig:uneven_repetitions}
    \end{subfigure}
    \caption{Comparison between UEP schemes with loss-based and Hessian-based sensitivity metrics ($\mathrm{SNR}_{train}=5\mathrm{dB}$, $b\approx 1.3\times 10^{6}$).}
    \label{fig:uneven_comparison}
\end{figure*}

Fig. \ref{fig:uneven_accuracy} compares different sensitivity measuring criteria (Hessian-based and loss-based) in terms of the accuracy they achieve. The number of repetitions per layer of the VGG16 network is presented in Fig. \ref{fig:uneven_repetitions}. In this comparison, we utilize the simple layer repetition scheme as it allows any integer number of repetitions, unlike the SK expansion method which is limited to per-layer repetition factor $r_i$ of the form $2^n$. We can observe that the Hessian-based strategy performs much better than the alternative schemes for every SNR value considered. Surprisingly, a large difference can be observed between the loss-based and Hessian-based sensitivity metrics in both the performance and the number of repetitions per layer. Despite being intuitively similar, the two methods prioritize different layers; the loss-based method prioritizes the later layers close to the output, while the Hessian-based method distributes the repetitions more evenly across the network. We can further notice that the layer repetition scheme with loss-based sensitivity is outperformed by the even repetition scheme in the low SNR regime, which illustrates that the protection of the early layers is more important than the later layers of the DNN.

\section{Conclusions}
\label{sec:conclusions}

In this work, we studied the important problem of the transmission of the DNN parameters over wireless channels, which is expected to become a significant traffic load for future networks given the increasing adoption of machine learning applications in edge devices. We have proposed training the DNN with noise injection to enable robustness against channel impairments, and network pruning to meet the bandwidth constraint. We have then shown that performance can be improved further, particularly in the low SNR regime by pruning the network to a size below the channel bandwidth and applying bandwidth expansion, which can be considered as an analog error correction technique. We then exploited the fact that not all DNN layers are equally significant for its final accuracy, and introduced an UEP technique by applying selective bandwidth expansion only to the most important layers of the network. Finally, to reduce the memory requirements caused by training a separate network targeting different channel conditions, we developed a novel ensemble training approach that allowed us to simultaneously train a whole spectrum of networks that can be adaptively used in different channel conditions. We believe this method lay the foundations for ``on demand'' delivery of DNNs in the future mobile networks.

\bibliographystyle{IEEEtran}
\bibliography{refs}

%







\end{document}